\documentclass[a4paper,11.5pt]{article}
\usepackage[top=30truemm,bottom=30truemm,left=30truemm,right=30truemm]{geometry}
\usepackage{hyperref}
\usepackage{amsmath}
\usepackage{ascmac}
\usepackage{amssymb}
\usepackage{ulem}
\usepackage{graphicx}

\begin{document}
\begin{flushright}
\end{flushright}
\begin{center}
{\LARGE Complexity of AdS$_5$ black holes with a rotating string}\\
\vspace{2cm}
{\large Koichi Nagasaki}\footnote{koichi.nagasaki24@gmail.com}\\
\vspace{5mm}
{\small Department of Physics and Center for High Energy Physics,
Chung Yuan Christian University\\
Address: 200 Chung Pei Road, Chung Li District, Taoyuan City, Taiwan, 32023}
\end{center}
\vspace{1cm}

\abstract{We consider computational complexity of AdS$_5$ black holes. 
Our system contains a particle moving on the boundary of AdS.
This corresponds to the insertion of a fundamental string in AdS$_5$ bulk spacetime.
Our results give a constraint for complexity. 
This gives us a hint for defining complexity in quantum field theories.}

\section{Introduction}
\subsection*{Background}
Computational complexity is originally introduced in physics of quantum information \cite{2008arXiv0804.3401W} and its application to black hole physics is considered in many works \cite{Roberts:2016hpo, Stanford:2014jda, Susskind:2014moa, Susskind:2014rva}. 
It behaves in a similar way to the entropy as it is expected to satisfy the second law of thermodynamics \cite{Brown:2017jil}.
The growth of complexity is related to problems of black holes, for example, information problem, the transparency of horizons or the existence of firewalls \cite{Almheiri:2012rt, Harlow:2013tf, Mann:2015luq, Susskind:2015toa, Polchinski:2016hrw}.

Complexity-Action (CA) conjecture \cite{Brown:2015bva, Brown:2015lvg} predicts that complexity of the black hole is equal to the the bulk action integrated in the region called the Wheeler-de Witt patch (WdW).
This conjecture is checked by many works \cite{Barbon:2015soa, Chapman:2016hwi, Carmi:2016wjl, Kim:2017lrw, Reynolds:2017lwq}.
The time dependence is studied in \cite{Pan:2016ecg, Brown:2016wib, Cai:2016xho, Alishahiha:2017hwg, Zhao:2017iul, Guo:2017rul, Wang:2017uiw}.
This duality is studied also in a deformed case by adding probe branes \cite{Abad:2017cgl}.

Nevertheless, the strict and valid definition of complexity is still unclear.
The most standard one so far is by the method of quantum operators which are called gates.
Let us define a reference state as the simplest quantum state.
Quantum gates operate a state and change its state.
Computational complexity of a test state is defined as the minimum number of gates to prepare it from the reference state.
But this definition has some unsatisfied points. 
One is that the definition of the reference state is unclear.
Another is that the number of the gates depends on the choices of the primitive gates.
 
\subsection*{Our approach}
To treat this problem, in this paper, we shed light on a new property of complexity and suggest a new constraint which complexity should possess.
The above works so far treat stationary systems. 
A generalization to this conjecture to the system other than stationary states is an important work.
As an example, we consider here a system moved by a drag force. 
Such a work is motivated by jet-quenching phenomena in heavy ion collisions.
Energy loss of a charged particle in quark-gluon plasma (QGP) is calculated in several approaches \cite{Gubser:2006bz, Herzog:2006gh, CasalderreySolana:2006rq, Liu:2006ug}.
Especially in \cite{Gubser:2006bz} a particle moving in AdS$_5$-Schewarzchild spacetime was considered and its action was calculated.
The particle is moved on the boundary of the AdS$_5$ space.
Because of the effect of shear viscosity in QGP, this particle losses the energy. 
This system is a dissipative system.

While his work considered the action in the outside of the horizon, that is usual spacetime, we apply here this method also to the inside of the horizon. 
According to CA relation, we will obtain complexity of a time dependent system.
In this work we consider a Wilson line operator located in AdS$_5$ spacetime by inserting a fundamental string. 
This Wilson line moves on a great circle in $S^3$ part in AdS$_5$.
Such a non-local operator describes a test particle moving on the boundary gauge theory.
This shows how complexity is deformed when adding a time-dependent operator, especially also non-local operator.
Since inserting the Wilson loop is described by adding a Nambu-Goto (NG) term, the action is expected to consist of the  Einstein-Hilbert term, the Nambu-Goto term and the boundary term.
We study on the effect of the Wilson line operator focusing on the Nambu-Goto term and show the black-hole mass and particle's velocity dependences of the action.
This will show the growth of complexity for a dissipative system.

This paper is constructed as follows.
In section \ref{Sec:setup}, we explain our setup which includes the AdS black hole with a kind of non-local operator, a Wilson line, and the action.
In section \ref{Sec:NGaction}, in order to find the effect of the test particle, we focus on the NG action which is a part of the action we introduced the former section.
Section \ref{Sec:discussion} gives an interpretation of our calculation and some suggestions of the constraint for complexity.

\section{Setup}\label{Sec:setup}
We consider a Wilson loop inserted on AdS$_5$ spacetime.
The black hole geometry in AdS$_5$ spacetime is described by the metric:
\begin{align}\label{eq:AdSmetric}
ds_\text{AdS}^2 &= -f(r)dt^2 + \frac{dr^2}{f(r)} + r^2d\Omega_3^2,\\
f(r) &= 1 - \frac{16\pi GM}{5\Omega_3r^2}
  + \frac{r^2}{\ell_\text{AdS}^2}
= 1 - \frac{r_\text{m}^2}{r^2} + \frac{r^2}{\ell_\text{AdS}^2},\qquad
r_\text{m}^2 := \frac{16\pi GM}{5\Omega_3},\label{def:shwarzrad}
\end{align}
where $d\Omega_3^2$ is the square of the line element on three sphere, while $d\Omega_3$ is the volume form on the three sphere and $\Omega_3$ is the volume obtained by integrating it.
\footnote{
The volume of $(n-1)$-dim unit sphere $S^{n-1}$ is $\Omega_{n-1} = \frac{2\pi^{n/2}}{\Gamma(n/2)}$.
}
We see in appendix \ref{sec;GTpara} that $\ell_\text{AdS}$ and other parameters are related to the gauge theory parameters as
\begin{align}\label{eq:defx}
\frac{r_\text{m}^2}{\ell_\text{AdS}^2} 
= \frac{16\pi GM}{5\Omega_3\ell_\text{AdS}^2}
= \frac{8GM}{5\pi\ell_\text{AdS}^2}
= \frac{4M}{5}\frac{\sqrt{\alpha'g_\text{YM}}}{N^{7/4}}.
\end{align}
This geometry has some characteristic length; one is the Schwarzschild radius defined in eq.\eqref{def:shwarzrad} and the other is $r_h$ which is the radius of the black-hole horizon which is determined by the equation $f(r) = 0$:
\begin{align}
r_h = \ell_\text{AdS}\Bigg(\frac{-1+\sqrt{1+4r_\text{m}^2/\ell_\text{AdS}^2}}{2}\Bigg)^{1/2}\label{eq:horizonradius}.
\end{align}
We parametrize $S^3$ part of the AdS spacetime \eqref{eq:AdSmetric} by 
\begin{align}
x^1 &= \cos\theta\cos\phi,\:\:
x^2 = \cos\theta\sin\phi,\nonumber\\
x^3 &= \sin\theta\cos\varphi,\:\:
x^4 = \sin\theta\sin\varphi;\\
&\theta\in[0,\pi/2),\:
\phi,\varphi\in[0,2\pi).\nonumber
\end{align}

Let us take parameters $\tau$ and $\sigma$ on the world-sheet of the fundamental string:
\begin{align}
t = \tau,\:\:
r = \sigma,\:\:
\phi = \omega\tau + \xi(\sigma),
\end{align}
where $\omega$ is the constant angular velocity and $\xi(\sigma)$ is a function which determines the shape of the string. 
Thus, the induced metric on the world-sheet is
\begin{align}
ds_\text{ind}^2 &= -f(\sigma)d\tau^2 + \frac{d\sigma^2}{f(\sigma)},\\
f(\sigma) &= 1 - \frac{r_\text{m}^2}{\sigma^2} + \frac{\sigma^2}{\ell_\text{AdS}^2}.
\end{align}
The action consists of the Einstein-Hilbert term and the Nambu-Goto (NG) term:
\begin{align}
S_\text{EH} 
&= \frac{1}{16\pi G}\int d^{6}x\sqrt{|g|}(\mathcal R-2\Lambda),\:\:
\Lambda = \frac{-6}{\ell_\text{AdS}^2},\\
S_\text{NG} 
&= -T_s\int d^{2}\sigma\sqrt{-\det g_\text{ind}},
\end{align}
where $T_s$ is the tension of the fundamental string.
We are afraid for the necessity of adding a boundary term like the York-Gibbos-Hawking (YGH) term.
The development of the Einstein-Hilbert action is given in \cite{Brown:2015lvg} as
\begin{align}
\frac{dS_\text{EH}}{dt}
&= -\frac{1}{2\pi G\ell_\text{AdS}^2}
  \int_\Sigma r^{3}dr d\Omega_{3}
= -\frac{1}{8\pi G}\frac{r_h^{4}\Omega_{3}}{\ell_\text{AdS}^2}.\label{eq:EHresult}
\end{align}
This is the leading contribution to complexity. In the next section we consider the effect of the addition of the Wilson loop.

\section{Evaluation of the NG action}\label{Sec:NGaction}
\begin{figure}[t] 
\begin{center}
\includegraphics[width = 12cm]{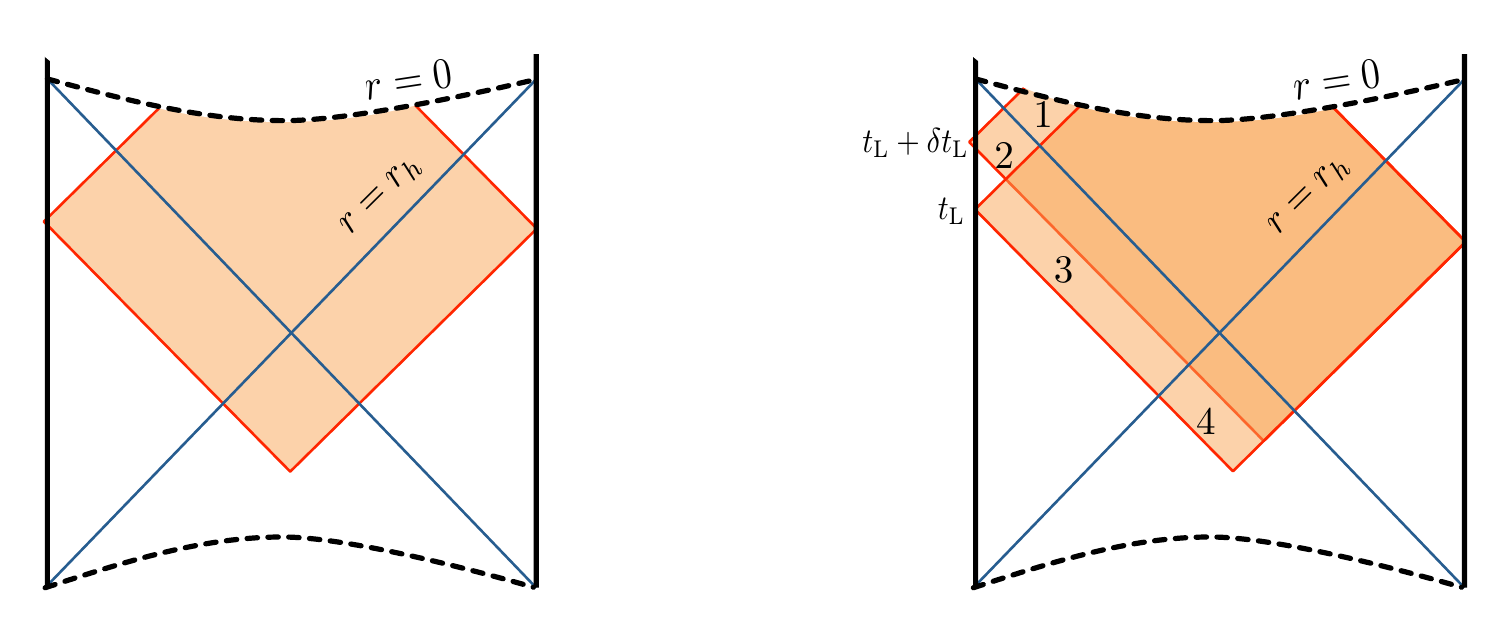}
\caption{Penrose diagram of the AdS black hole. The blue diagonal lines represent the black hole horizon.
(left) 
The shaded rectangular region represents the Wheeler-de Witt patch (WdW).
(right) 
Two WdW in different left CFT time $t_L$.
The regions 1,2,3 and 4 represent the difference of these two pathces.
The contributions from regions 2 and 3 cancel.
The contribution from region 4 becomes smaller as time develops. Then the integration over only the region 1 is relevant for our analysis.
}\label{fig:WdWpatch}
\end{center}
\end{figure}  
The addition of the Wilson loop corresponds to inserting a fundamental string whose world-sheet has a boundary on the Wilson loop.
We calculate here the NG action of this fundamental string.
The time derivative of the NG action is obtained by integrating the square root of the determinant of the induced metric over the WdW (figure \ref{fig:WdWpatch}):
\begin{align}\label{eq:tdepWdWaction}
\frac{dS_\text{NG}}{dt} 
&= T_s\int_{0}^{r_h} d\sigma
	\sqrt{1 - \frac{\sigma^2\omega^2}{f(\sigma)} + \sigma^2\xi'(\sigma)^2f(\sigma)}.
\end{align}
The Lagrangian is
\begin{align}
\mathcal L
= T_s\sqrt{1 - \frac{\sigma^2\omega^2}{f(\sigma)} + \sigma^2\xi'(\sigma)^2f(\sigma)}.
\end{align}
Then the equation of motion for $\xi(\sigma)$ is 
\begin{align}
0 = \frac{1}{T_s}\frac{d}{d\sigma}\frac{\partial\mathcal L}{\partial\xi'(\sigma)}
&= \frac{d}{d\sigma}\Bigg(\frac{\sigma^2 \xi'(\sigma)f(\sigma)}{\sqrt{1- \sigma^2\omega^2/f(\sigma) + \sigma^2 \xi'(\sigma)^2 f(\sigma)}}\Bigg),\\
c_\xi 
&:=  \frac{\sigma^2 \xi'(\sigma)f(\sigma)}{\sqrt{1- \sigma^2\omega^2/f(\sigma) + \sigma^2 \xi'(\sigma)^2 f(\sigma)}}
=  \frac{\sigma^2 \xi'(\sigma)f(\sigma)}{\mathcal L/T_s},
\end{align}
where by the second line we defined a conserved constant $c_\xi$.
Solving it for $\xi(\sigma)$, this function is given by integrating the following:
\begin{align}\label{eq:xieq1}
\xi'(\sigma) 
= \frac{c_\xi}{\sigma f(\sigma)} 
	\sqrt\frac{f(\sigma) - \sigma^2\omega^2}{\sigma^2f(\sigma) - c_\xi^2}.
\end{align}
In order for this expression to give real values, the denominator in the square root must be negative when the numerator factor $f(\sigma) - \sigma^2\omega^2$ becomes negative. 
Thus they become zero coincidentally.
From this condition, the constant $c_\xi$ is determined to be 
\begin{align}
c_\xi = \omega\sigma_H^2
&= \omega\frac{-1+\sqrt{1+4r_\text{m}^2(1/\ell_\text{AdS}^2-\omega^2)}}{2(1/\ell_\text{AdS}^2-\omega^2)},\\
\sigma_H^2 &= \frac{-1+\sqrt{1+4r_\text{m}^2(1/\ell_\text{AdS}^2-\omega^2)}}{2(1/\ell_\text{AdS}^2-\omega^2)}\label{def:sigmaH},
\end{align}
where $\sigma = \sigma_H$ is the solution for numerator of the square root in eq.\eqref{eq:xieq1}.
We assumed that $c_\xi$ is positive.
Since the numerator and the denominator have the coincident solution, the cancelation gives
\begin{align}
\xi'(\sigma) 
&= \frac{\omega\sigma_H^2}{(\sigma^2 - r_h^2)(1/\ell_\text{AdS}^2(\sigma^2 + r_h^2) 
+ 1)} 
  \sqrt\frac{(1/\ell_\text{AdS}^2-\omega^2)(\sigma^2 + \sigma_H^2) 
+ 1}{1/\ell_\text{AdS}^2(\sigma^2 + \sigma_H^2) + 1},\label{eq:xidiff}\\
\mathcal L 
&= T_s\sigma^2\xi'(\sigma)f(\sigma)/c_\xi
= T_s\sqrt\frac{(1/\ell_\text{AdS}^2-\omega^2)(\sigma^2 + \sigma_H^2) 
+ 1}{1/\ell_\text{AdS}^2(\sigma^2 + \sigma_H^2) + 1}\label{eq:laglangian1}.
\end{align}
The integral of \eqref{eq:laglangian1} over the WdW is 
\begin{align}
\frac{dS_\text{NG}}{dt} 
&= \int_\text{WdW} \mathcal L
= T_s\int_{0}^{r_h} d\sigma
	\sqrt\frac{(1/\ell_\text{AdS}^2-\omega^2)(\sigma^2 + \sigma_H^2) + 1}{1/\ell_\text{AdS}^2(\sigma^2 + \sigma_H^2) + 1}.
\end{align}
Using the integral formula \eqref{eq:intform1} shown in appedix,
\begin{align}
\frac{dS_\text{NG}}{dt} 
= -iT_s\sqrt{\sigma_H^2(1-\ell_\text{AdS}^2\omega^2)+\ell_\text{AdS}^2}
  E\Bigg[\arcsin\Big(i\frac{r_h}{\sqrt{\sigma_H^2 + \ell_\text{AdS}^2}}\Big),
    \sqrt\frac{(\sigma_H^2 + \ell_\text{AdS}^2)(1-\ell_\text{AdS}^2\omega^2)}{\sigma_H^2(1-\ell_\text{AdS}^2\omega^2) + \ell_\text{AdS}^2}\:\Bigg].
\end{align}
Substituting eq.\eqref{def:sigmaH} and the horizon radius $r_h$, \eqref{eq:horizonradius}, we can express the growth of the action as
\begin{align}
\frac{dS_\text{NG}}{dt} 
&= -iT_s\ell_\text{AdS}
 \Bigg(\frac{1+\sqrt{1+4s^2
    (1-v^2)}}{2}\Bigg)^{1/2}\times\nonumber\\
&\qquad
E\Bigg[\arcsin\Big(i\Big(
  \frac{(-1+\sqrt{4s^2+1})(1-v^2)}{(1 - 2v^2) +\sqrt{1+4s^2(1-v^2)}}\Big)^{1/2}\Big),
    \Big(\frac{(1 - 2v^2) + \sqrt{1+4s^2(1-v^2)}}  
  {1+\sqrt{1+4s^2(1-v^2)}}\Big)^{1/2}\:
  \Bigg],
\end{align} 
where for simplicity we defined the following dimensionless constants:
\begin{align}
s := r_\text{m}/\ell_\text{AdS}, \:\:
v := \ell_\text{AdS}\omega.
\end{align}

\begin{figure}[t] 
\begin{center}
\includegraphics[width = 10cm]{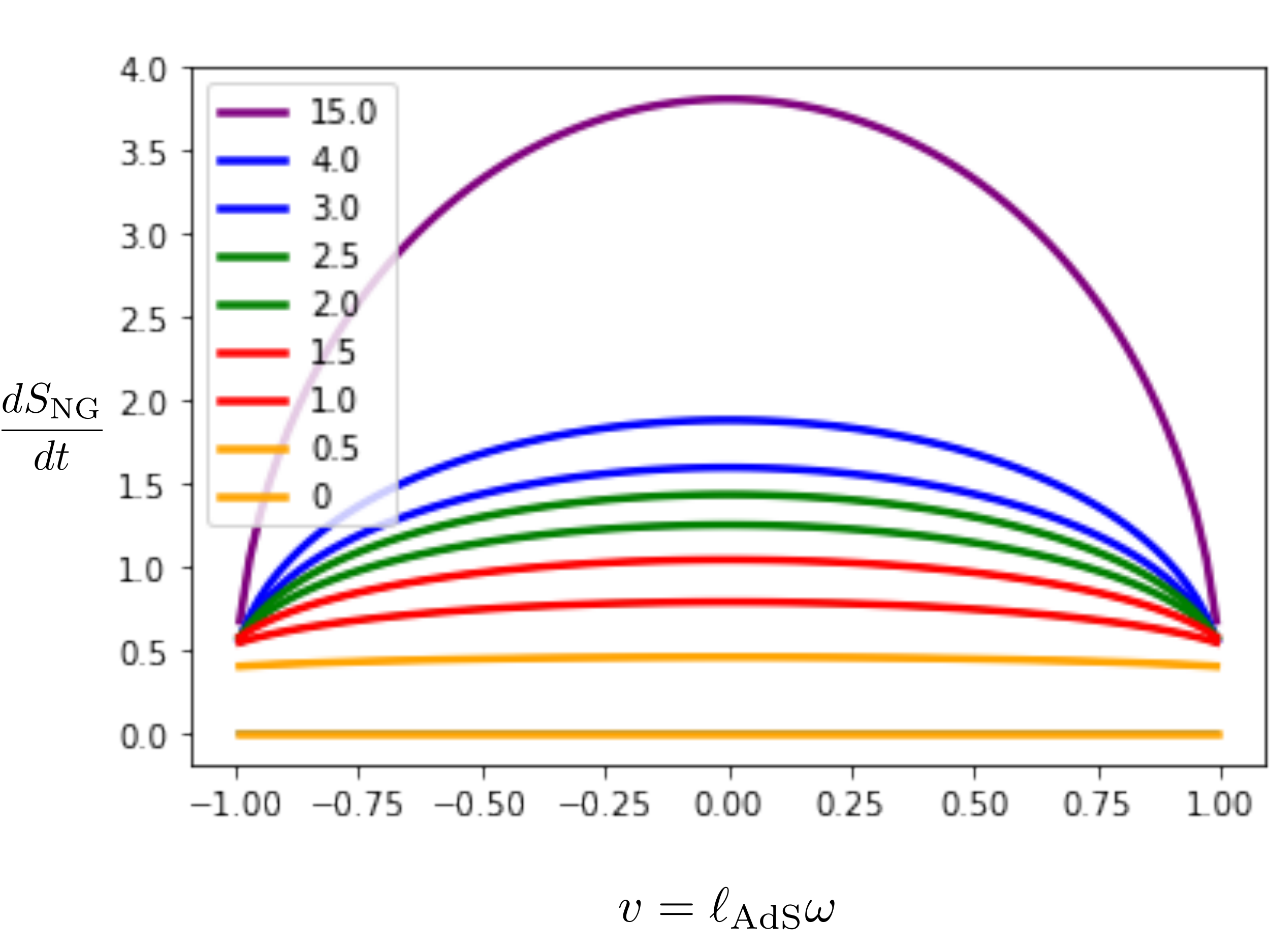}
\caption{Angular velocity dependence}\label{fig:actionvelocityplot}
\end{center}
\end{figure}

\begin{figure}[t] 
\begin{center}
\includegraphics[width = 10cm]{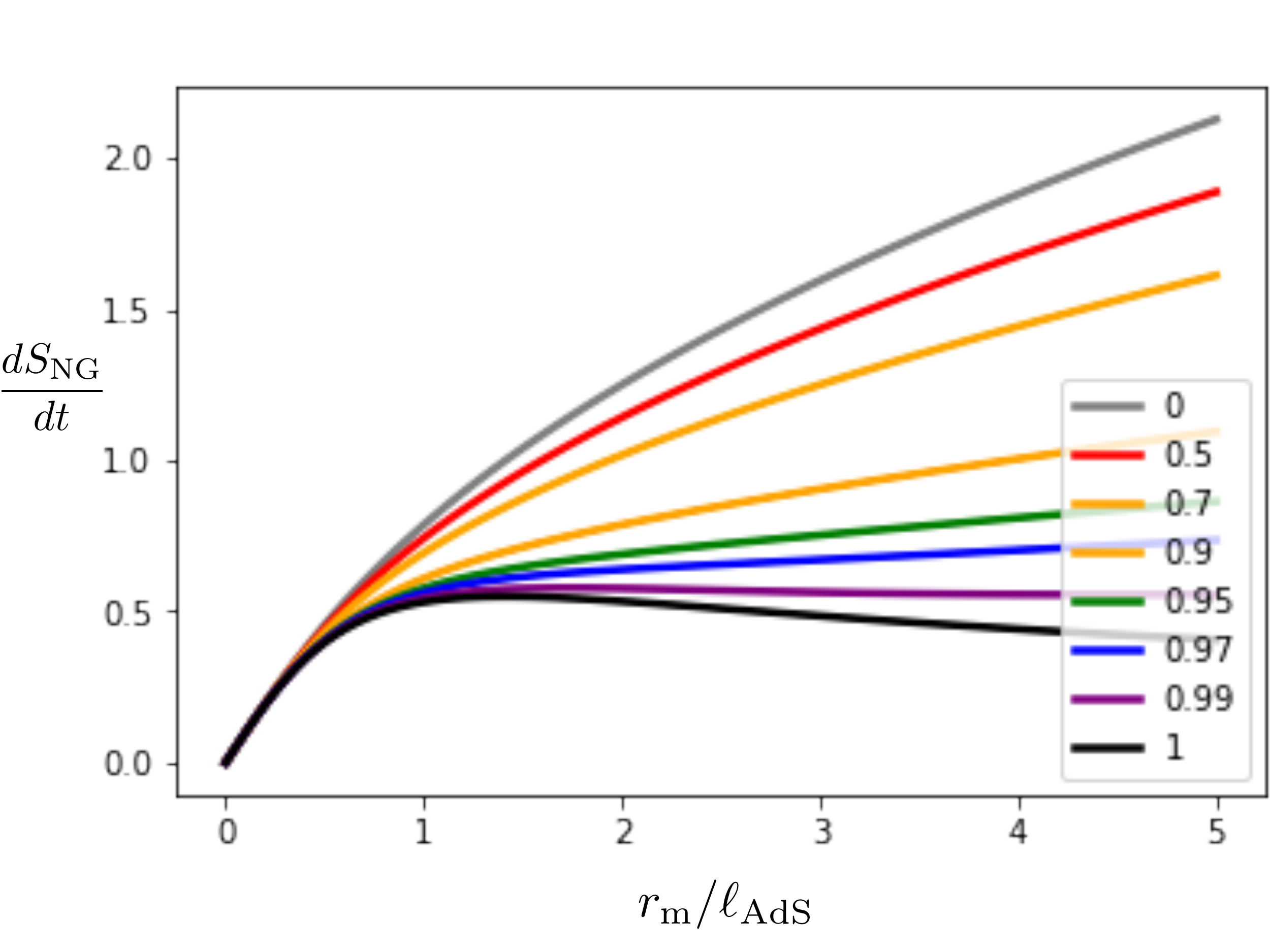}
\caption{Mass dependence}\label{fig:actionmassplot}
\end{center}
\end{figure}

\paragraph{Limit behavior} 
The behavior in the limits $s\rightarrow 0$ and $s\rightarrow\infty$ are, respectively, as follows:
\begin{align}
\frac{dS_\text{NG}}{dt}\Big|_{s\rightarrow 0}
&= -i4T_s\ell_\text{AdS} E\Big[0, \sqrt{1-v^2}\Big] = 0,\\
\frac{dS_\text{NG}}{dt}\Big|_{s\rightarrow\infty}
&= 4T_s\ell_\text{AdS}\sqrt{s}(1-v^2)^{1/2}.
\end{align}
The behavior in the limits $v\rightarrow 0$ and $v\rightarrow 1$ are 
\begin{align}
\frac{dS_\text{NG}}{dt}\Big|_{v=0}
&= \sqrt{2}T_s\ell_\text{AdS}
  \Big(\frac{-1+\sqrt{4s^2+1}}{2}\Big)^{1/2} \sim \sqrt s \:\:(s\gg 1),\\
\frac{dS_\text{NG}}{dt}\Big|_{v\rightarrow 1}
&= -iT_s\ell_\text{AdS}
  \arcsin i\Big(\frac{-1+\sqrt{4s^2+1}}{2(1+s^2)}\Big)^{1/2}
= T_s\ell_\text{AdS}
  \text{arcsinh}\Big(\frac{-1+\sqrt{4s^2+1}}{2(1+s^2)}\Big)^{1/2}\label{eq:limitv1}.
\end{align}
The figures \ref{fig:actionvelocityplot} and \ref{fig:actionmassplot} show these behavior and also the curves of all parameter ranges.
The function \eqref{eq:limitv1} has an extremum at $s=\sqrt 2$.

\section{Discussion}\label{Sec:discussion}
Figure \ref{fig:actionvelocityplot} shows the relation between the growth of the action and the angular velocity ($v = \ell_\text{AdS}\omega$). 
We can see that the effect of the drag force to the growth of the action is larger as the mass increases when the string moves relatively slower. As the string moves faster the effect become smaller.
For any mass, the effect vanishes in the limit $v\rightarrow 1$.
This behavior is similar to the one that of a rotating black holes --- the complexity growth bound changes from $2M$ to $2\sqrt{M^2 - J^2/\ell_\text{AdS}^2}$, where $J$ is the angular momentum of the black holes.

Figure \ref{fig:actionmassplot} shows mass dependence of the growth of the action. 
When the angular velocity is small, it is a monotonically increasing function of the mass.
As the angular velocity becomes larger, it ceases to increase rapidly and the extremum appears at about $v\gtrsim 0.97$.
That is a notable phenomena --- complexity has different velocity dependence in the relativistic region.

According to CA duality these represent the growth of the black-hole complexity.
For a large mass black hole, complexity changes rapidly when the particle moves slowly and this effect becomes small for a relativistic particle.
As the black hole becomes smaller, the complexity does not change whether the particle moves quickly or not.
Complexity is expected to increase as the entropy increases by the second law of thermodynamics \cite{Brown:2017jil}.
So the results shown in figure \ref{fig:actionvelocityplot} and figure \ref{fig:actionmassplot} are consistent with this expectation.

The velocity dependence of energy loss is changed by particle's mass.
It is due to collisions between quarks and gluons, and gluon bremsstrahlung and which is dominant depends on its velocity \cite{Herzog:2006gh}.
The behavior of the action for large mass, $(\sim\sqrt{s})$, is consistent with the expected behavior from CA relation \cite{Brown:2015lvg}, that is, the growth of complexity is bounded by two times mass. 
These behavior characterize the growth of complexity and gives a hint to define complexity in black hole spacetime.

In our analysis the boundary term, which was detaily discussed in \cite{Carmi:2016wjl}, is not taken into consideration.
An example of the boundary term calculation is given in \cite{Drukker:1999zq}.
They considered AdS boundaries.
It is different from our case since in our calculation we need boundary terms which come from the boundary of region 1 (see figure \ref{fig:WdWpatch}). 
There is a cancelation and the contributions only from $r=0$ (black-hole singularity) and $r=r_h$ (horizon) survive.
The detailed analysis will be considered as a future work.
However, while the Einstein-Hilbert term includes the second derivative of the metric, the NG action only includes the first derivative of it.
The value at the boundary is completely determined by the metric and the derivative of it is unnecessary.
Then we can expect that the boundary term is not need in this situation.

\section*{Acknowledgments}
I would like to thank Satoshi Yamaguchi, Hiroaki Nakajima and Sung Soo Kim for useful discussions.
This work is supported in parts by the Chung Yuan Christian University, the Taiwan's Ministry of
Science and Technology (grant No. 105-2811-M-033-011)

\appendix
\section{Gauge theory parameter}\label{sec;GTpara}
The Newton's gravitational constant we used is related to the gauge theory parameter.
Since 5-dimensional gravitational constant and AdS radius are as follows \cite{Maldacena:1997re}:
\begin{align}
&G_5 = \frac{\pi\ell_\text{AdS}^3}{2N^2},\:\:
\ell_\text{AdS}^2 = \alpha'\sqrt{g_\text{YM}^2N}
\:\:\Rightarrow\:\:
G_5 
= \frac{\pi}{2N^2}\alpha'^{3/2}(g_\text{YM}^2N)^{3/4}
= \frac{\pi}{2}\frac{(\alpha' g_\text{YM})^{3/2}}{N^{5/4}},
\end{align}
we obtain
\begin{align}
G_5/\ell_\text{AdS}^2 
= \frac{\pi}{2N^2}\sqrt{\alpha'}(g_\text{YM}^2N)^{1/4}
= \frac{\pi\sqrt{\alpha'g_\text{YM}}}{2N^{7/4}}.
\end{align}
Note that some people may use the different convention where $\ell_\text{AdS}^2$ is different by the factor $\sqrt 2$, for example \cite{johnson2006d}, but we follow the above definition in this paper.

\section{Black hole temperature}\label{sec:BHtemp}
\begin{figure}[h] 
\begin{center}
\includegraphics[width = 9cm]{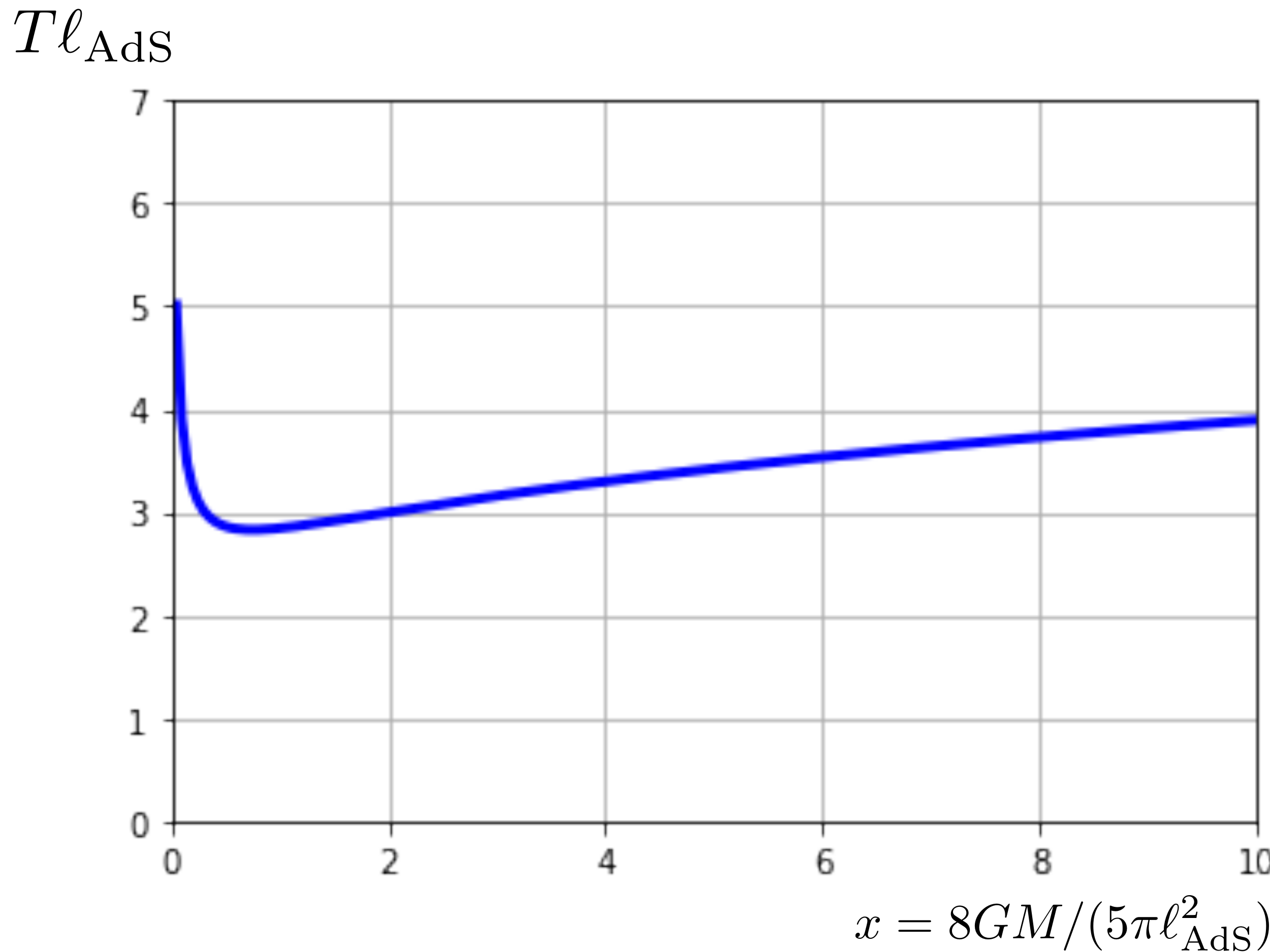}
\caption{Temperature - Mass}\label{fig:tempmass}
\end{center}
\end{figure}

Here we derive the temperature of AdS$_5$ black hole.
This result can be seen in \cite{Witten:1998zw}.
We consider the Euclidean version of the metric \eqref{eq:AdSmetric} and take the coordinates:
\begin{align}
r = r_h(1+\rho^2).
\end{align}
Near the horizon the metric is approximated as
\begin{align}
ds_\text{AdS}^2
&\approx 
2\rho^2\Big(\frac{r_\text{m}^2}{r_h^2} + \frac{r_h^2}{\ell_\text{AdS}^2}\Big) dt_\text{E}^2
 + 2r_h^2\Big(\frac{r_\text{m}^2}{r_h^2} + \frac{r_h^2}{\ell_\text{AdS}^2}\Big)^{-1}d\rho^2
 + r_h^2d\Omega_3^2\nonumber\\
&= 2r_h^2\Big(\frac{r_\text{m}^2}{r_h^2} + \frac{r_h^2}{\ell_\text{AdS}^2}\Big)^{-1}
  \Big(d\rho^2
 + \frac{\rho^2}{r_h^2}
    \Big(\frac{r_\text{m}^2}{r_h^2} + \frac{r_h^2}{\ell_\text{AdS}^2}\Big)^2 dt_\text{E}^2\Big)
 + r_h^2d\Omega_3^2.
\end{align}
Seeing the period of Euclidean time $t_\text{E}$, the black hole temperature ($1/\beta$) is 
\begin{align}
\beta 
&= 2\pi r_h
  \Big(\frac{r_\text{m}^2}{r_h^2} + \frac{r_h^2}{\ell_\text{AdS}^2}\Big)^{-1}.
\end{align}
Expressed it by using mass, this leads 
\begin{align}
T
&= \frac{1}{2\pi r_h}
  \Big(\frac{r_\text{m}^2/\ell_\text{AdS}^2}{r_h^2/\ell_\text{AdS}^2} 
  + \frac{r_h^2}{\ell_\text{AdS}^2}\Big)\nonumber\\
&= \frac{1}{2\pi\ell_\text{AdS}}
  \Big(\frac{-1+\sqrt{1+4x}}{2}\Big)^{-1/2}
  \Big(\frac{2x}{-1+\sqrt{1+4x}} 
  + \frac{-1+\sqrt{1+4x}}{2}\Big),\:\:
x := \frac{8GM}{5\pi\ell_\text{AdS}^2},\nonumber\\
T\ell_\text{AdS}
&= \frac{1}{2\pi}
  \Big(x\Big(\frac{-1+\sqrt{1+4x}}{2}\Big)^{-3/2}
  + \Big(\frac{-1+\sqrt{1+4x}}{2}\Big)^{1/2}\Big).
\end{align}
where we used eqs. \eqref{eq:horizonradius} and \eqref{eq:defx} and in the last line we used dimensionless quantities $T\ell_\text{AdS}$ and $x$.
This function is plotted in figure \ref{fig:tempmass}.

\section{Elliptic Integral}\label{sec:ellipticint}
For calculating the action in the Wheeer-de Witt patch (WdW) we use the elliptic integral of second kind.
The elliptic integral of the second kind is defined as 
\begin{align}\label{def:2ndelliptic}
E[\varphi, k]
:= \int_{0}^{\sin\varphi}\sqrt\frac{1-k^2t^2}{1-t^2}\:dt.
\end{align}
We need the following integral formula:
\begin{align}
\int_{0}^{c}\sqrt\frac{x^2+a}{x^2+b}dx
= -i\sqrt a E\Big[\arcsin\Big(i\frac{c}{\sqrt b}\Big),\sqrt\frac{b}{a}\:\:\Big].\label{eq:intform1}
\end{align}
It can be proved as
\begin{align}
\int_{0}^{c}\sqrt\frac{x^2+a}{x^2+b}dx
&= \sqrt\frac{a}{b}\int_{0}^{c}\sqrt\frac{1 + x^2/a}{1 + x^2/b}dx\nonumber\\
&=\sqrt\frac{a}{b}\int_{0}^{ic/\sqrt{b}}\sqrt\frac{1 - (b/a)t^2}{1 - t^2}(-i\sqrt{b}dt),\:\:
	ix/\sqrt{b} = t,\nonumber\\
&= -i\sqrt{a} E\big(\arcsin(ic/\sqrt{b}, \sqrt{b/a})\big).
\end{align}

\providecommand{\href}[2]{#2}\begingroup\raggedright\endgroup

\end{document}